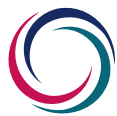

# Computational tools for drawing, building and displaying carbohydrates: a visual guide

Kanhaya Lal[‡1,2], Rafael Bermeo[‡1,2] and Serge Perez[*1]

**Review**

Address:
[1]Univ. Grenoble Alpes, CNRS, CERMAV, 38000 Grenoble, France and [2]Dipartimento di Chimica, Università Degli Studi di Milano, via Golgi 19, I-20133, Italy

Email:
Serge Perez[*] - spsergeperez@gmail.com

[*] Corresponding author    ‡ Equal contributors

## Abstract

Drawing and visualisation of molecular structures are some of the most common tasks carried out in structural glycobiology, typically using various software. In this perspective article, we outline developments in the computational tools for the sketching, visualisation and modelling of glycans. The article also provides details on the standard representation of glycans, and glycoconjugates, which helps the communication of structure details within the scientific community. We highlight the comparative analysis of the available tools which could help researchers to perform various tasks related to structure representation and model building of glycans. These tools can be useful for glycobiologists or any researcher looking for a ready to use, simple program for the sketching or building of glycans.

## Introduction

Glycoscience is a rapidly surfacing and evolving scientific discipline. One of its current challenges is to keep up and adapt to the increasing levels of data available in the present scientific environment. Indeed, the rise of accessible experiment data has changed the landscape of how research is performed. The accessibility of this information, coupled with the emergence of new platforms and technologies, has benefitted glycoscience to the point of enabling the detection and high-resolution determination and representation of complex glycans [1]. Increasing numbers of carbohydrate sequences have accumulated throughout extensive work in areas of chemical and biochemical fragmentations followed by analysis using mass spectroscopy, nuclear magnetic resonance, crystallography and computational modelling. There have been some initiatives by independent research groups worldwide, that pushed the development of visual tools to improve some aspects of glycan identification, quantification and visualisation, some of which will be further developed throughout this article.

Biological molecules express their function throughout their three-dimensional structures. For this reason, structural biology places great emphasis on the three-dimensional structure as a central element in the characterisation of biological function. An adequate understanding of biomolecular mechanisms inherently requires our ability to model and visualise them. Visualisation of molecular structures is thus one of the most common tasks performed by structural biologists. As an essential part of the research process, data visualisation allows not only to communicate experimental results but also is a crucial step in the integration of multiple data derived resources, such as thermodynamics and kinetic analysis, glycan arrays, mutagenesis, etc. Data visualisation remains a challenge in glycoscience for both the developers and the end-users even for the simple task of describing molecular structures. Progress in this area allows to translate a static visualisation of single molecules into dynamic views of complex interacting large macromolecular assemblies, which increases our understanding of biological processes.

Representing the structures of carbohydrates has historically been considered to be a complicated task. Starting from the linear form of the Fischer projection, which is certainly not a realistic representation of a carbohydrate structure, there has been a continuous development and evolution of the description of monosaccharides [2]. Glycans are puzzles to many chemists, and biologists as well as bioinformaticians. This complexity occurs at different levels (which makes it incremental). Amongst the most recognisable "sugars", glucose is merely one of 60+ monosaccharides, all of which are, in truth, pairs of mirror-image enantiomers (ʟ and ᴅ).

Moreover, monosaccharides occur as two forms: 5-atom ring (furanose) and 6-atom ring (pyranose). With the occurrence of a statistically rarer "open form," we obtain at least 6 "correct" representations of glucose. And yet, monosaccharides are only the chemical units and the individual building blocks of much more complex molecules; the carbohydrates, also referred to as glycans. The glycan family can be grouped in the following categories: (i) oligosaccharides (comprising two to ten monosaccharides linked together either linearly or branched); (ii) polysaccharides (for glycan chains composed of more than ten monosaccharides); (iii) glycoconjugates (where the glycan chains are covalently linked to proteins (glycoproteins), lipids (glycolipids). The complexity of glycans is a consequence of their branched structure and the range of building blocks available. Other levels of complexity include the nature of the glycosidic linkage (anomeric configuration, position and angles), the number of repeating units (polysaccharides) as well as the substitutions of the monosaccharides. Regardless of the different nomenclatures available to describe each monosaccharide, representing and encoding a glycan structure into a file is required for communication among scientists as well as for data processing.

As a consequence, glycobiologists have proposed different graphical representations, with symbols or chemical structures replacing monosaccharides. The description of carbohydrate structures using standard symbolic nomenclature enables easy understanding and communication within the scientific community. Research groups working on carbohydrates have developed schematic depictions with symbols [3] and expansions with greyscale colouring as the so-called Oxford nomenclature (UOXF) [4,5], and even fully coloured schemes later on. Among these, some of the proposed representation forms have been accepted and implemented by several groups and initiatives, namely the Consortium for Functional Glycomics (CFG) [6]. Whereas the initial versions of such representation were limited to mammalian glycans, an extension of the graphical representation of glycans, called SNFG Symbol Nomenclature for Glycans (SNFG) [7,8] resulted from a joint international agreement. The newly proposed nomenclature covers 67 monosaccharides aptly represented in eleven shapes and ten colours. There is the hope that it will cope better with the rapidly growing information on the structure and functions of glycans and polysaccharides from microbes, plants and algae. The rendering of glycan drawing and symbol representations motivated the development of several computer applications using a standardised notation. The earliest glycan editors allowed manual drawing similar to ChemDraw or used input files with glycan sequence KCF (KEGG Chemical Function) [9] in text format for similarity search against other structures deposited in the databases. Later developments supported the construction and representation of glycan structures in symbolic form by computational tools like GlycanBuilder [10]. Since then, several advancements have been made to allow the user to both draw glycans manually or by importing and exporting the structure files in different text formats [11].

Along the same line, the development of various other applications allowed the users to sketch 2D-glycan structures by dragging and dropping monosaccharides to canvas to generate 3D structures for further usages. These depictions comply with protein data bank (PDB) [12] format, or in the form of images [13,14]. Besides, these tools for representing glycans in 2D and 3D shape [15] allowed the integration of glycans into protein structures or complexes. The tools developed in the last few years have automated the sketching of glycans and glycopeptides, allowing rapid display of structures using IUPAC format [16] as input. This article explores and illustrates the concepts of "sketching", "building" and "viewing" glycans (Figure 1). It provides a descriptive analysis of the tools available for such activities, which can be useful for researchers looking for a ready-to-use simple program for sketching, building and 3D structure analysis of glycans and glycoconjugates. The scope of this work is relevant to N- and O-linked glycans, glycolipids, proteoglycans and glycosaminoglycans, lipopolysaccharides, plant, algal and bacterial polysaccharides.

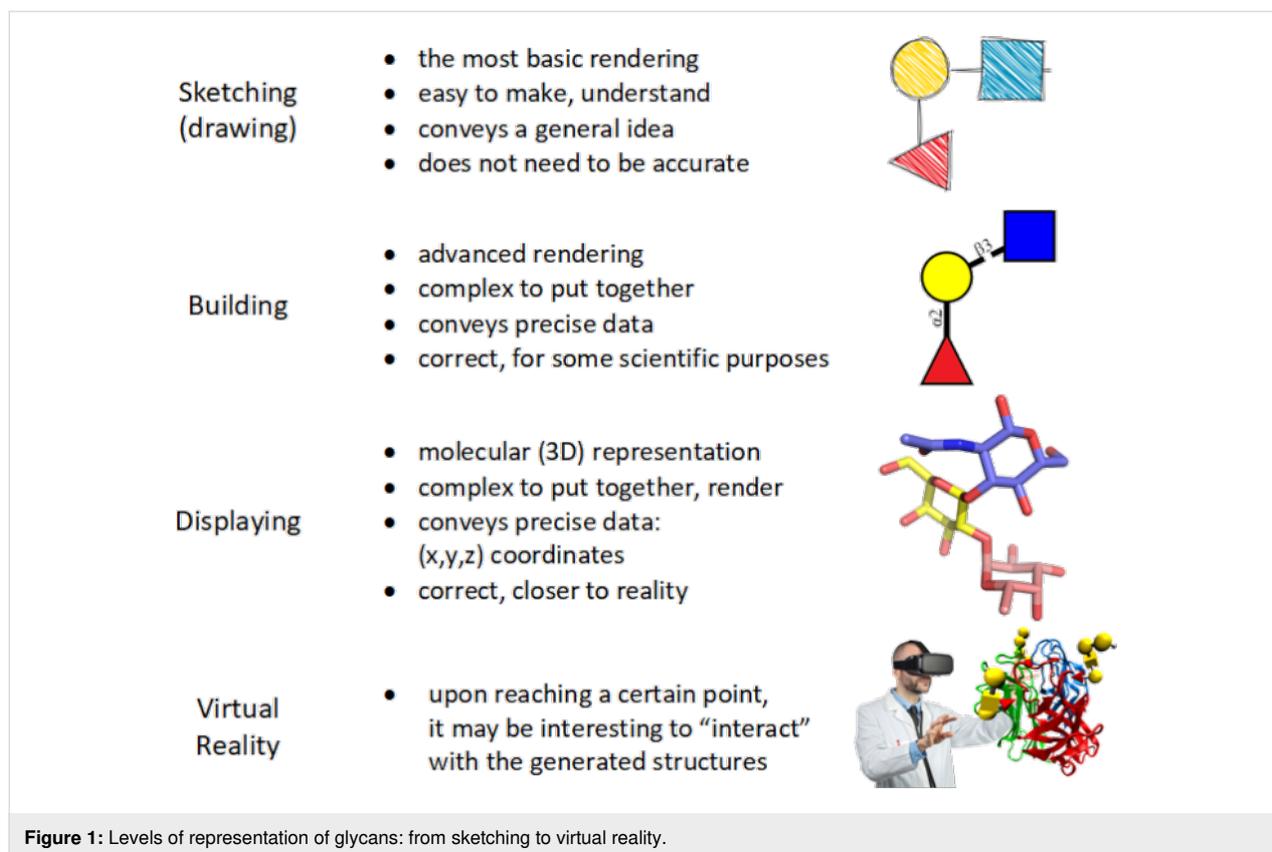

**Figure 1:** Levels of representation of glycans: from sketching to virtual reality.

## Review
### Methods

To facilitate glycoscience research, we have identified the tools and databases that are freely available on the internet and are regularly updated and improved [1]. The variety and complexity of glycan structures make their interpretation challenging. Consequently, in the past few years, several sketching, building and visualisation tools have been developed to depict better and understand the complex glycan structures. In this study, the freely available tools have been visited (April 2020) and analysed to highlight their core features but also explore their unique advancements to facilitate glycan research. Each of the computational tools was inspected for general features related to sketching, representing and model building, all of which could be further used as input for translation into other formats, search from glycan databases or complex calculations such as molecular simulations. Several tools feature an interactive interface which allows for manual editing of the structures. Examples of such tools are DrawRINGS [17], KegDraw [18], Glycano (available at http://glycano.cs.uct.ac.za/), GlycoEditor [19], GlycanBuilder [20], etc. These tools (except KegDraw) are provided with the list of CFG symbols to freely build glycan structures using the mouse on the canvas. In addition to manual sketching, some of these tools also can import text formats including IUPAC-condensed, GlycoCT and KEGG Chemical Function (KCF) format to display the glycan structures. Some applications also facilitate glycan search in various databases. Another category of tools included in this study involved glycan viewers which can only depict structures using the IUPAC three letter code or IUPAC-condensed nomenclature as input. These tools convert the input into a 2D image or 3D representation using SNFG symbols or 3D-SNFG illustration. Additionally, 3D representation of structures is provided by tools such as Visual Molecular Dynamics (VMD) [21] , and LiteMol [22], which allow for quick analysis of structural features in 3D space. All the tools mentioned were evaluated against a set of pre-selected criteria relating to ease of use, scientific precision and content, among others.

Table S1 (Supporting Information File 1) schematically summarises how these criteria are fulfilled. The analysis of the tools for input and output formats also provided information about their versatility to convert results into the standard or desired format. The tools have been attributed to categories such as "sketcher", "builder" and "viewer", with eventual overlaps. A brief analysis of each application ordered by category is given in the next section.

## Sketching with the free hand

As a preview of the following parts of this study, we performed an initial test of the tools available for the representation of a simple disaccharide: lactose (β-D-Gal*p*-(1→4)-D-Glc*p*).

Figure 2 shows how different web-available platforms rendered it. On the one hand, thanks to the unified nomenclature, there is no ambiguity regarding the nature of the carbohydrate represented. On the other hand, small differences between sketches appear. Such variations will multiply with the increasing complexity of the carbohydrates. It is, therefore, essential to choose which tools to use before starting an hour-long "drawing-spree". The variations of the colour code used to represent the monosaccharides show striking differences across platforms even though the appropriate colours to be used are strictly defined (https://www.ncbi.nlm.nih.gov/glycans/snfg.html#tab2). The colour discrepancy observed here means that some of the tools do not conform to SNFG standards. For some purposes, this conformity might not be a strict necessity. Another pronounced disparity concerns the representations of the glycosidic linkage. Across sketches the length/width of the linkage varies, which will result in either compact or extended images, to be taken into account when considering the size available for the intended figures. Finally, the sketches provide further information about the linkage type: anomericity and position. These details can be either useful or superfluous depending on what is the intended use for the finished design. The main characteristic of a helpful sketching tool should be its adaptability. By allowing to modify colours, sizes, lengths/widths and turn some features on/off, a "sketcher" would allow maximum flexibility to depict carbohydrates in any desired or necessary form, size, orientation. However, this adaptability should become available without hampering the sketching effort. The perfect sketching tool would, therefore, combine flexibility and high usability.

## Building with scientific accuracy

The necessity for precision is what, at some point, turns carbohydrate sketching into building. What defines this turning point (besides a certain level of accuracy) is the intended purpose for the produced figures/images. Scientific communication, comparison between similar yet different structures, or merely showcasing the complexity of carbohydrates: all three cases cannot rely on a sketching tool to convey their message. Consequently, a new set of considerations appears. The requirement for accurate depiction comes from the complexity mentioned above of carbohydrates: anomeric configuration, substitution, glycosidic bond position, and repeating units (as well as tethering to larger macromolecules, and more). For the sake of accuracy, only the right combination of characteristics should be depicted, leaving no ambiguity: every relevant piece of data should be detailed. The glycosidic linkage is a perfect example to illustrate the necessity for accuracy in building, as opposed to sketching. While a simple line is enough to link two monosaccharides, it is necessary to define the linkage as alpha or beta (or unknown) and to state the positions of the glycosyl acceptor and even donor. Cellulose and amylose are two glucose-based polysaccharides that differ only in the nature of their glycosidic bond, and yet they have entirely different shapes and so, biological roles. For the sake of completion, the full description of a monosaccharide should obey the following rules: *<anomeric prefix><prefix for absolute configuration><the monosaccharide code><prefix for ring configuration>[<O-ester and O-ether substitutions and positions>]*. It is thus necessary to include such information when depicting carbohydrates, but such features are simply absent in most of the existing glycan sketching tools.

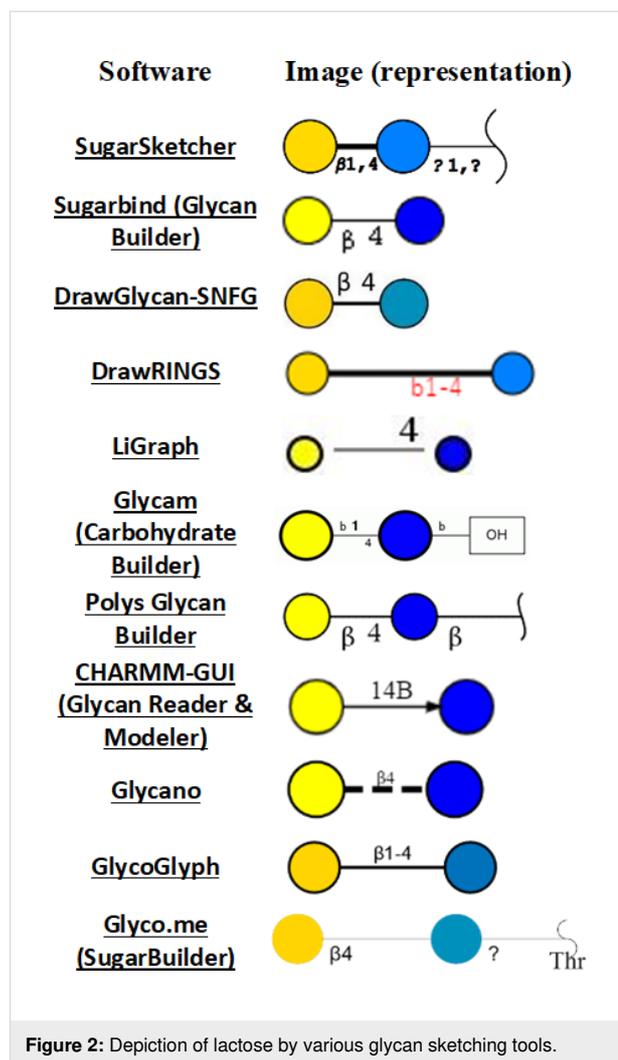

**Figure 2:** Depiction of lactose by various glycan sketching tools.

Another feature that may become essential when the carbohydrate at hand is a polysaccharide is the possibility of building repeating units. Without this option, it would be simply impossible to build the required depiction. It emerges that an efficient carbohydrate builder must offer a wide array of options to characterise and personalise each monosaccharide. This would, in turn, entail a multitude of buttons, switches, etc.; which would result in a very complex interface. Consequently, unless the interface is rather straightforward and the building dynamic is well-designed, the software would be too difficult to use effectively. The ideal carbohydrate builder pick would also allow liberty for the user in terms of levels of precision since it has to fit every level of complexity above sketching. Lastly, once the building process is complete, a good builder must not only render all the provided data in the form of a precise figure but also allow the transfer of the data to other platforms (for example, by exporting the generated code).

## Force fields for carbohydrates, 3D model building and beyond

Carbohydrates present various challenges to the development of force fields [23]. The tertiary structures of monosaccharides usually have a high number of chiral centres which increases the structural diversity and complexity. The structural diversity changes the electrostatic landscape of molecules; thus, it provides challenges in the development of force fields for accurate modelling of such variations in charge distributions. The monosaccharides can further form a large number of oligosaccharides which can enormously increase the conformational space, due to a high number of rotatable bonds. Nonetheless, recent developments in carbohydrate force fields enable to model and reproduce the energies associated with minute geometrical changes. The currently available force fields which are parameterised for carbohydrates are also capable of carrying out simulations of the oligosaccharides containing additional groups like sulfates, phosphates etc. [24] Generally used force fields for the Molecular Mechanics (MM) simulation of carbohydrates are CHARMM [25], GLYCAM [26], and GROMOS [27]. The structural complexity increases the computational cost, which makes simulations of large systems more challenging. Therefore, coarse-grained models [28] for carbohydrates are generally used for molecular modelling of large systems.

In terms of 3D model building, the complex topologies of glycans require dedicated molecular building procedures to convert sequence information into reliable 3D models. These tools generally use 3D molecular templates of monosaccharides to reconstruct a 3D model. Energy minimisation methods can further refine the models. These models are essential for structure-based studies and complex calculations like Molecular Dynamics simulations. Therefore, the accurate model building requires the use of reliable databases to generate atomic coordinates and topology to provide an acceptable model. Some of the computational tools usually contain atom coordinates of generally used monosaccharides (as templates) and also use libraries of bond and angle parameters from various force fields dedicated for carbohydrates. The accurately predicted oligosaccharide conformations are good starting points for further investigations. Of particular interest are the evaluations of the dynamics of glycans and their interactions with proteins which is a most significant concern in glycoscience. The joint need to better perceive and manipulate the three-dimensional objects that make up molecular structures is leading to a rapid appropriation of techniques of Virtual Reality (VR) by the molecular biology community. Generic definitions describe VR as being immersion in an interactive virtual reactive world. The computer-generated graphics provide a realistic rendering of an immersive and dynamic environment that responds to the user's requests. One finds in these definitions the three pillars that define VR: Immersion, Interaction, Information. Although it is difficult to extract a single, simple definition of VR, the main idea is to put the user at the centre of a dynamic and reactive VR environment, artificially created and which will supplant the real world for the time of the experiment.

## Input and output for sketching, building and displaying applications

The variety and complexity of carbohydrate structures hamper the use of a unique nomenclature. The choice of notation depends on whether the study is focused on chemistry or has a more biological approach. The IUPAC-IUBMB (International Union for Pure and Applied Chemistry and International Union for Biochemistry and Molecular Biology) terminologies, in their extended and condensed forms [16], govern the naming of the primary structure or sequence.

Further down the line, the complexity of the existing nomenclatures for carbohydrate-containing molecules remains a significant hurdle to their practical use and exchanges within and outside the glycoscience cenacle. The linearisation of the description of the structure is a way to cope with the description of the structural complexity. The proposed formats provide rules to extract the structure of the branches and create a unique sequence for the carbohydrate. The most commonly used formats are IUPAC [16], GlycoCT [29], KCF [9], and WURCS [30].

The sketching of carbohydrates using computational tools generally requires the textual input and output in at least one of

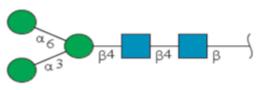

| Input Output formats | |
|---|---|
| IUPAC condensed | Man(a1-3)[Man(a1-6)]Man(b1-4)GlcNAc(b1-4)b-GlcNAc |
| LINUCS | []{b-D-GlcPNAc}{[(4+1)][b-D-GlcPNAc]{[(4+1)][b-D-Manp]{[(3+1)][a-D-Manp]{}[(6+1)][a-D-Manp]{}}}} |
| GlycoCT | RES<br>1b:b-dglc-HEX-1:5<br>2s:n-acetyl<br>3b:b-dglc-HEX-1:5<br>4s:n-acetyl<br>5b:b-dman-HEX-1:5<br>6b:a-dman-HEX-1:5<br>7b:a-dman-HEX-1:5<br><br>LIN<br>1:1d(2+1)2n<br>2:1o(4+1)3d<br>3:3d(2+1)4n<br>4:3o(4+1)5d<br>5:5o(3+1)6d<br>6:5o(6+1)7d |
| KCF | ENTRY     G12157          Glycan<br>NODE      6               EDGE      5<br>     1 Asn 18 0        1 2:b1 1:4<br>     2 GlcNAc 9 0      2 3:b1 2:4<br>     3 GlcNAc -1 0     3 4:b1 3:4<br>     4 Man -11 0       4 5:a1 4:6<br>     5 Man -19 3       5 6:a1 4:3<br>     6 Man -19 -3      /// |
| WURCS | WURCS=2.0/3,5,4/[a2122h-1b_1-5_2*NCC/3=O][a1122h-1b_1-5]/1-1-2-3-3/a4-b1_b4-c1_c3-d1_c6-e1 |

these formats (Figure 3). An alternate input method involves manual sketching of 2D glycan structures by dragging and dropping monosaccharide symbols on canvas (with or without grids) to connect them further. This method makes the sketching tools more friendly and interactive as it does not require large text code as input. Both input methods are compliant to the Symbol Nomenclature for Glycans (SNFG). Another symbolic representation that could clearly distinguish monosaccharides in monochrome colours is the Oxford notation [5]. In this method, dashed and solid lines represent the alpha and beta glycosidic linkages, respectively. There are few tools which have implemented this method while other tools use text to highlight this information in the structures. In addition to sketching tools, some applications, specific to the field of carbohydrates, provide the possibility to visualise and display 3D structures. These visualisation tools accept strings or files in text formats (GlycoCT, IUPAC-condensed, KCF) to display the structure via a graphical user interface. For instance, the DrawGlycan-SNFG [31] tool uses IUPAC-condensed nomenclature for input string and converts it into a 2D image represented in SNFG symbols. At the same time, the 3D-SNFG [15] can generate glycan structures by incorporating SNFG symbols in 3D space for further visualisation using the computational tools like visual molecular dynamics (VMD) [21] LiteMol [22] and Sweet Unity Mol [32].

## Glycan sketchers

**SugarSketcher.** SugarSketcher [14] is a JavaScript interface module currently included in the tool collection of Glycomics@ExPASy (available at https://glycoproteome.expasy.org/sugarsketcher/) for online drawing of glycan structures. The interactive graphical interface (Figure 4, top) allows glycan drawing by glycobiologists and non-expert users. In particular, a "Quick Mode" helps users with limited knowledge of glycans to build up a structure quickly as compared to the normal mode, which offers options related to the structural features of complex carbohydrates (for example additional monosaccharides, isomers, ring types, etc.). The building of glycan structures uses mouse and proceeds via a selection of monosaccharides, substituents and linkages from the list of symbols. However, some wrong combinations of choices can block the interface, resulting in the need to re-start the process (SugarSketcher is on version beta 1.3). Alternatively, SugarSketcher also uses GlycoCT or a native template library as an input. A list of pre-built core N- and O-linked carbohydrate moieties, which are usually present in glycoproteins structures, can be used as a template for further modification. A shortlist of glycan epitopes is also included providing templates for drawing more complex molecules. The software uses the Symbol Nomenclature for Glycans (SNFG) notation for structure representation and exports the obtained sketch to text format (GlycoCT) or image (.svg) files. The software SugarSketcher is featured in the web portal GlyCosmos (https://glycosmos.org/glytoucans/graphic) [33]. Under the name "SugarDrawer", it provides an interface for generating carbohydrate structures to query the database included in GlyCosmos: GlyTouCan [34].

GlyCosmos is a web portal that integrates resources linking glycosciences with life sciences. Besides elements such as "SugarDrawer" and GlyTouCan (carbohydrate database), the platform GlyCosmos assembles data resources ranging from glycoscience standard ontologies to pathologies associated with glycans. GlyCosmos is recognized as the official portal of the Japanese Society for Carbohydrate Research and provides information about genes, proteins, lipids, pathways and diseases.

GlyTouCan (Figure 5) is a repository for glycans which is freely available for the registry of glycan structures. The repository can register structures ranging from monosaccharide compositions to fully defined structures of glycans. It assigns a unique accession number to any glycan to identify its structure and even allows to know its ID number in other databases. Al-ternatively, users can search and retrieve information about the glycan structures and motifs that have been already registered into the repository. The structures can be searched simply by browsing through the list of already registered glycans or by specifying a particular sub-structure to retrieve structurally similar glycans (https://glytoucan.org/Structures/graphical). The software tool featured in the GlyTouCan website is called GlycanBuilder and is presented in a later section of our analysis.

**Figure 4:** From top to bottom: SugarSketcher [36] interface with a glycan structure drawn using the "Quick Mode". LiGraph interface showing input and output options for glycan structure representation. GlycoGlyph [37] interface with a text input (modified IUPAC condensed) converted into its glycan image.

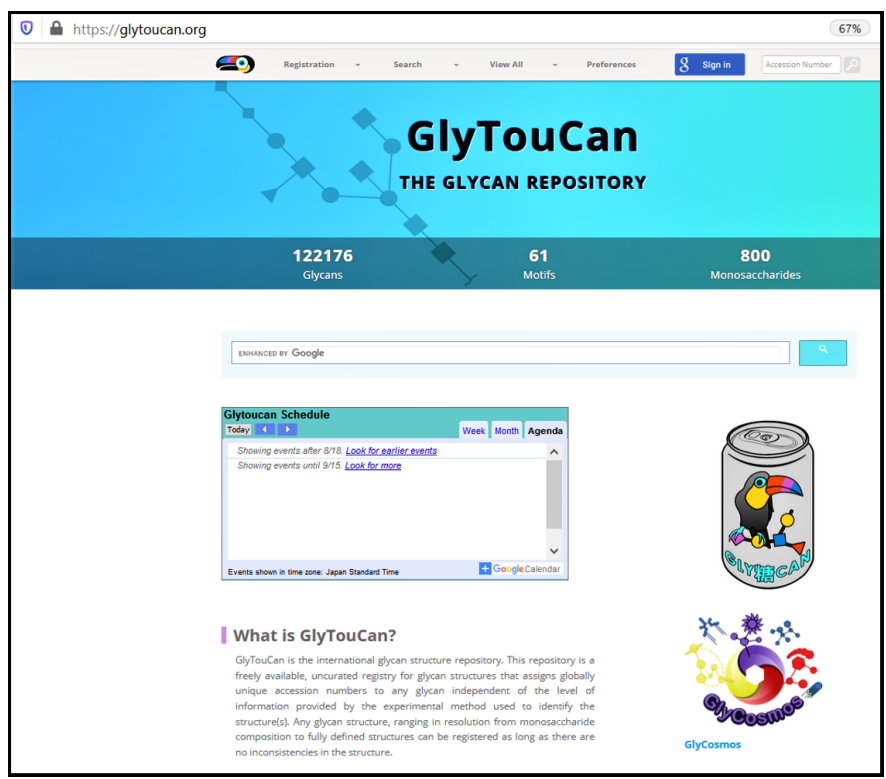

**Figure 5:** GlyTouCan [38] interface allows to search for glycans structures in the database. Data contained in GlyCosmos portal (https://glycosmos.org/) and in GlyTouCan repository home page (https://glytoucan.org/), including their logos, are licensed under a Creative Commons Attribution 4.0 International License (https://creativecommons.org/licenses/by/4.0/).

Recapitulating, SugarSketcher can be an efficient tool for non-glycobiologists or glycobiologists to sketch glycans. However, it does not accept different input or output formats like IUPAC, WURCS (Web3 Unique Representation of Carbohydrate Structures), which would make the tool more versatile.

**LiGraph.** LiGraph [35] (http://www.glycosciences.de/tools/LiGraph/) is an online tool based on the concept of schematic drawings of oligosaccharides to display glycan structures. This tool also renders images of glycans in different notation using a text input. The input for the carbohydrate structure consists of a list of names and connections. The glycan structure is output in the specified notation: either ASCII IUPAC sugar nomencla-ture or a graph which can be rendered in different themes which include Heidelberg, Oxford, Tokyo, CFG and extended CFG (Figure 4, middle). The output images for the glycan structure and the legends can be saved and downloaded in .svg format. This tool is useful for glycan sketching using text templates, but its shortcomings include a limited number of monosaccharide symbols and restricted compatibility with other input file formats.

**GlycoGlyph.** GlycoGlyph [39] is a web-based application (available at https://glycotoolkit.com/Tools/GlycoGlyph/) built using JavaScript which allows users to draw structures using a graphical user interface or via text string in the CFG linear (also known as modified IUPAC condensed) nomenclature dynamically. The interface (Figure 4, bottom) is equipped with templates for N- and O-linked glycans and terminals. Also, it provides 80+ monosaccharide (SNFG) symbols and a selection for substituents. The selected template or text string (in CFG linear nomenclature) input directly gets converted into an image in canvas and also appears as text in GlycoCT format. The output can be saved as a .svg file or as GlycoCT text. The interface also provides additional options to add, replace or delete each monosaccharide, modify the sizes of symbols and text fonts, and turn off the linkage annotations or change their orientation; all of which increases the usability of the software. The input structure can be further used to search the GlyTouCan [34] database to explore the literature details related to the input structure.

GlycoGlyph is an efficient tool for sketching or building glycans with a highly usable interface that can significantly help researchers to improve the uniformity in glycan formats in literature/manuscripts. It can also be a tool of choice for text mining for the query structure.

**GlycanBuilder2.** GlycanBuilder2 [40] is a Java-based glycan drawing tool which runs locally as an application on different platforms including Windows, macOS and Linux. It is freely available for downloading at http://www.rings.t.soka.ac.jp/downloads.html. GlycanBuilder2 is a newer version of Glycan-Builder [20] with additional features. This version is capable of supporting various ambiguous glycans consisting of monosaccharides from plants and bacteria. The tool uses the SNFG notation to display glycan structures. Moreover, this updated version can convert a drawn structure into WURCS sequences for further use as a query for glycan search or registration in databases like GlyTouCan. GlycanBuilder2 provides an excellent interface (Figure 6, top) for glycan drawing. Glycan structures can be drawn manually using the mouse or by importing text input files. The interface provides a list of templates: N- O-glycans, glycosphingolipids, glycosaminoglycans(GAGs). Rows of CFG notations for monosaccharides assist with glycan structure drawing on canvas. The application also supports the glycan symbol notations for the University of Oxford (UOXF) format. The input complies with various linear sequence and text formats. They include GlycoCT, GLYcan structural Data Exchange using Connection Tables (GLYDE-II), Bacterial Carbohydrate Structures DataBase (BCSDB) [41], carbohydrate sequence markup language (CabosML) [42], CarbBank [43], LinearCode [44], LINUCS, IUPAC-condensed and Glyco-suiteDB [45]. The output yields structures in the following formats: GlycoCT, LinearCode, GLYDE-II and LINUCS. Thus, GlycanBuilder2 is a versatile tool which can be used for glycan sketching or building and also as a glycan sequence converter from one format to another.

**Original GlycanBuilder.** GlycanBuilder [10,20] was originally part of the GlycoWorkbench platform [49]. This interface is integrated in most tools of the Glycomics@ExPASy collection that require a drawing interface to query data. Glycan-Builder is written in Java Programming language and can be used as standalone or as an applet for embedding in web pages for glycan search. For example, GlycanBuilder is integrated in SugarbindDB [50] to draw glycan structures and search the database (https://sugarbind.expasy.org/builder), and in GlycoDigest or GlyS3 [10,20] to define the input of these tools.

Technically, the tool provides an interactive interface which allows an automated glycan rendering using a library of individual monosaccharides or pre-built template structures (Figure 6, middle). GlycanBuilder provides access to 41 templates. They include N- and O-linked glycans, GAGs (glycosaminoglycans), glycosphingolipids and milk oligosaccharides. It also contains 68 entries from MonosaccharideDB (http://www.monosaccharidedb.org/) including monosaccharides, modifications (e.g. deoxy) and substituents. The tool provides options to modify a monosaccharide by adding substituents and alterations. Free movement of the monosaccharides is allowed through movement and orientation buttons. Glycan-Builder offers multiple options for glycan notation which include CFG, CFG colour, UOXF, UOXF colour and text only. GlycanBuilder can also calculate the masses of glycan structures according to the options selected by the user. Glycan-Builder is a versatile tool for building carbohydrates, with multiple options for exporting the generated structures in the form of text format (GlycoCT, LINUCS, Glycominds, Glyde II) or image (.svg, .png, .jpg, .bmp, .pdf, etc.) files.

**DrawRINGS.** DrawRINGS [17] is a Java-based applet for rendering glycan structures on canvas (http://www.rings.t.soka.ac.jp/drawRINGS-js/). The different drawing features in an interactive interface (Figure 6, bottom) can be selected with the mouse by surfing the buttons and scroll-down menus. Alternatively, KCF files or KCF text format can be used as input. The free movement of the monosaccharides allows drawings with flexible geometry, for example, for schematic studies of carbohydrates. The drawn glycan structure can be exported in the KCF or IUPAC text format or saved in .png format. The drawn structure can further be used as a query for the search in glycan databases; using match percentage (Similarity) or by the number of components matched (Matched) criteria. Four predefined score matrices are available, named: N-glycans, O-glycans, Sphingolipids and Link_similarity. The "Link_similarity" matrix is based on glycosidic linkages and monosaccharides that may be more highly substituted with other glycosidic linkages and monosaccharides, respectively. There is a query to search the generated structure in the RINGS or GlycomeDB databases (or both). The former compiles data from the KEGG GLYCAN and GLYCOSCIENCES.de databases. DrawRINGS is an efficient tool for sketching glycan figures as well as translating to (and from) the KCF and IUPAC text formats.

**DrawGlycan-SNFG.** DrawGlycan-SNFG [31] is an open-source program available with a web interface (Figure 7, top) at

**Figure 6:** From top to bottom: GlycanBuilder2 [46] interface with a glycan image in SNFG notation. Original GlycanBuilder [47] interface with some of the available templates rendered as images. DrawRINGS [48] interface featuring a glycan and its KCF text output.

**Figure 7:** From top to bottom: DrawGlycan-SNFG [51] web interface with a glycan text input and the resulting image output. Glycano [52] interface with a glycan structure. GlycoEditor [53] interface, linkage selection is triggered by adding a new monosaccharide.

http://www.virtualglycome.org/DrawGlycan. The same web page gives access to a downloadable, standalone Graphical User Interface (GUI) version of this tool with additional functionality. It can be launched from different platforms including Windows, Mac or Linux. The program can be used to render glycans and glycopeptides using SNFG and uses IUPAC-condensed text inputs. The DrawGlycan-SNFG version with command-line operations makes it more versatile as it allows integration of multiple features of the program using custom scripts. The tool uses automatic operations for the majority of the drawing, which could meet the needs of researchers, but additional intervention may sometimes be required to get the desired output. For example, manual input in IUPAC-condensed language allows to generate, among others: repeating units, adducts, tethering to other structures (represented by text), and complex branching (the examples section showcases these options). The drawn glycan structure can be saved as .jpg image and modified through parameters such as symbol and text size, the thickness of lines, orientation of drawing and spacing. This software provides all the guidance and tools needed to generate high-quality pictures. DrawGlycan-SNFG is a reliable choice for building glycans.

In addition to glycan structure drawing, DrawGlycan-SNFG (version 2) [54] is equipped with a wide range of options to enhance the usability of the original code [32]. The new version is capable to accommodate the latest updates to the SNFG [7]. This tool has been particularly upgraded for MS spectrum annotation by adding an intuitive interface with additional features. The upgraded version can depict bond fragmentation, repeating structural unit anomeric groups, adduct ions, different types of glycosidic linkages etc. These advanced features make this tool ideal for integrated use with various glycoinformatics software and also for applications in glycoproteomics, glycomics and mass spectrometry (MS). One of the illustrations involves combined use with the gpAnnotate application, dedicated to score and annotate MS/MS glycopeptide spectrums in different fragmentation modes [54].

**Glycano.** Glycano (available at http://glycano.cs.uct.ac.za) is a software tool for drawing glycans. This tool is based on JavaScript, which can be used without the requirement of any server or browser dependency. The interactive interface allows sketching via the drag-and-drop method on canvas (with or without grid). The software is provided with "UCT" and "ESN", interchangeable interfaces (Figure 7, middle) with different symbols for monosaccharides. These names (UCT and ESN) correspond to the University of Cape Town, South Africa, where Glycano was developed, and to the "Essentials of Glycobiology Symbol Nomenclature", precursor of the SNFG symbol set [55]. The interface provides a wide choice of monosaccha-rides and substituents represented in SNFG symbols but lacks the standard colour scheme. The user can easily modify the structure with by click and drag, which allows to either cut/ copy, delete or move a portion of the structure. The drawn structure can be saved in text format, in .gly format or as an image (PNG and SVG formats). A drawback to note is that linking the monosaccharides at specific positions is only possible in the UCT mode, which means that back-and-forth between the two symbol systems is necessary to define the linkages correctly. Despite some drawbacks, this is an excellent tool due to its ease-of-use, tenable degree of freedom, and functionalities/options for sketching and building glycan structures.

**GlycoEditor.** GlycoEditor [19] (available at https://jcggdb.jp/ idb/flash/GlycoEditor.jsp) is an online software for drawing glycans. Through a straightforward interface, three ways of input are possible: by JCGGDB ID, through a library of common oligosaccharides and by direct input. A list of most common monosaccharides is presented, and the rest can be found categorised by family. The click and drag addition of new monosaccharides trigger the selection of linkage-type and configuration (Figure 7, bottom). The tool provides an option to create repeating units. Additionally, several functionalisation options are also available. Once the structure is ready, the user can save it as an .xml file. GlycoEditor allows searching a given structure across many databases in four ways: exact structure match (with or without anomer and linkage specifics) and the same for substructure match. The central database featured is the JCGGDB, to which can be added, among others: Glaxy, GlycomeDB, GlycoEpitope, GMDB, KEGG, etc. Searching by ID is also possible. GlycoEditor is a now dated tool that allows efficiently building glycans and performing databases searches.

**GLYCO.ME (SugarBuilder).** Glyco.me-SugarBuilder (available at https://beta.glyco.me/sugarbuilder) is online software for drawing glycans. The interface leads to rapid carbohydrate construction. A panel of monosaccharide templates complements the drawing interface (one pre-built oligosaccharide is available (Figure 8, top). The user can start a chain from amino acid residues: Asn, Ser or Thr, then structure building is limited by to a set of "rules" (limiting building options to known carbohydrates). These rules may be deactivated with a switch button to draw freely. A list of 13 monosaccharides is deployed, and sequential clicking allows their addition to the existing structure and definition of the associated glycosidic bond (the relative sizes of the options available related to their real statistical value for that particular linkage). Upon building some specific motifs, if they are recognised, an option for repeating units appears. Other switch buttons allow the user to change the orientation of the drawing, show/hide linkage information etc. The Oxford notation can be enabled for glycosidic bonds only. The structure obtained can be rendered as .png or .svg images. Glyco.me-SugarBuilder is still under development: more monosaccharides/substitutions/templates will complete an already very functional platform. The quick and easy options put forward offer natural building and liberty for tailoring the rendered image.

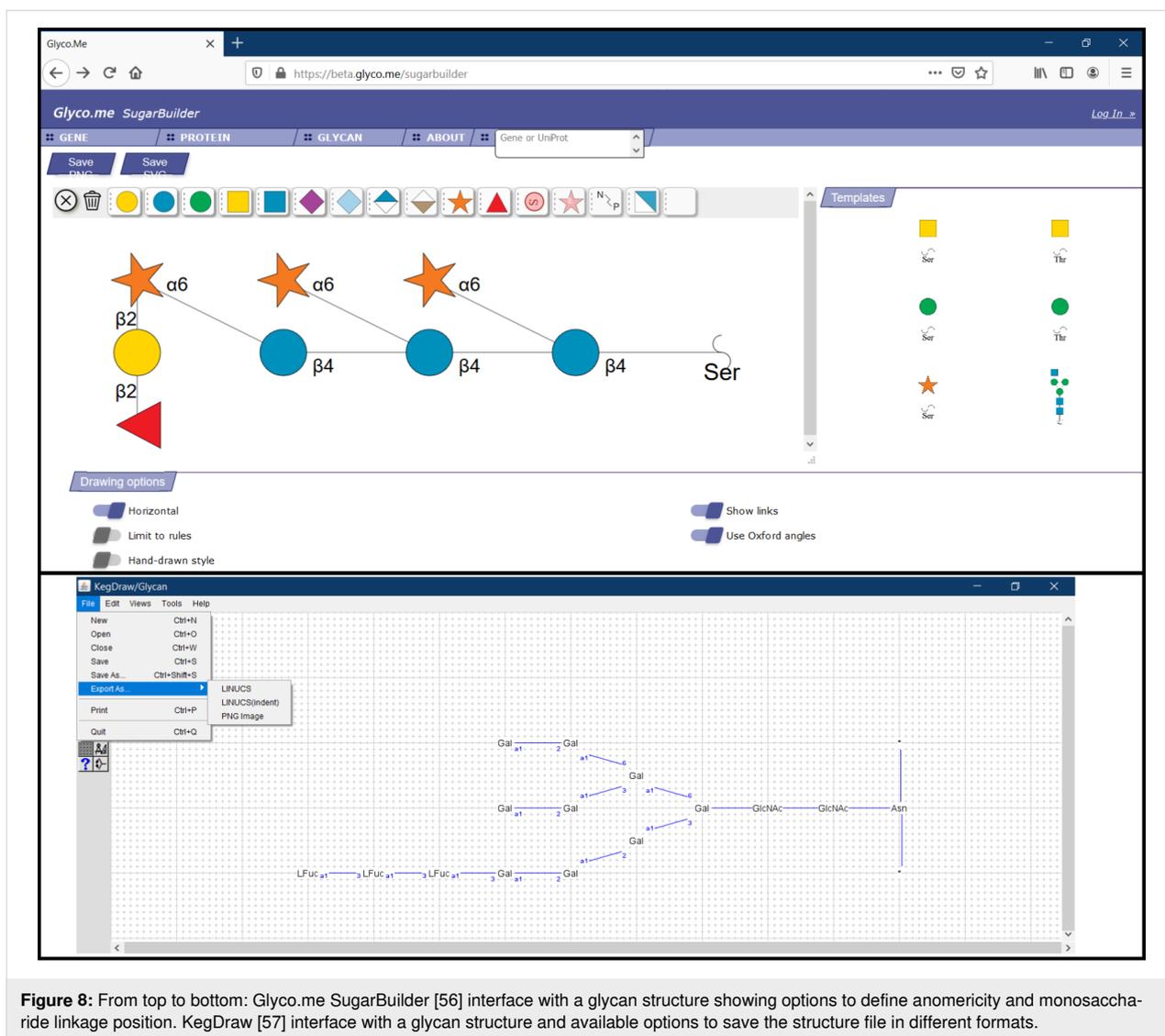

**Figure 8:** From top to bottom: Glyco.me SugarBuilder [56] interface with a glycan structure showing options to define anomericity and monosaccharide linkage position. KegDraw [57] interface with a glycan structure and available options to save the structure file in different formats.

**KegDraw.** KegDraw (https://www.kegg.jp/kegg/download/kegtools.html) is a freely available Java application for rendering glycan structures. It can be downloaded and installed locally as a platform-independent tool. This tool can be used in two different modes: "Compound mode" which can be used for drawing small molecules (similarly to any chemical structure drawing software), and "Glycan mode" which is dedicated to rendering glycan structures using different monosaccharide units. The simplest method for drawing involves a selection of monosaccharides and glycosidic linkages from an available list to generate a glycan structure. Alternatively, a text box option provides a way to draw uncommon types of monosaccharides. The tool also contains templates from KEGG GLYCAN and their importation using their accession number. Besides, input files in KCF can be used while the output can be saved in LINUCS, KCF or an image in PNG format (Figure 8, bottom). The glycan structure in text format can be further used as a query for search in KEGG GLYCAN and CarbBank databases. Hence, KegDraw can be an option for the freely available tool for drawing and querying chemical structures. However, there are similar tools already available for glycan drawing with more advanced and acceptable notations.

## Glycan builders
**Sweet II.** Sweet [58] is a web-based program for constructing 3D models of glycans from a sequence using standard nomenclature accessible at http://www.glycosciences.de/modeling/

sweet2/doc/index.php (Figure 9, top). This tool is available as a part of the glycosciences.de website, which also provides other options for analysing glycans in three-dimensional space. This program uses a glycan sequence in a standard format and generates a 3D model in the form of a .pdb file. The glycan input can come from a library of relevant oligosaccharides, available through one of the sub-menus. Alternatively, manual input is possible in three platforms adapted for increasing complexity. The model can be further minimised using MM2 [59] and MM3 [60] methods. The 3D models can be viewed using molecular viewers like JMol, WebMol-applet, Chemis3D-applet, etc. Besides, the program also generates additional files which can be used for molecular mechanics and molecular dynamics using molecular modelling tool like Tinker [61]. This tool is as a versatile tool for generating a 3D model for glycans.

**GLYCAM-web (Carbohydrate Builder).** Carbohydrate builder [65] is an online tool (at http://glycam.org/) for carbohydrate structure drawing and subsequent 3D structure building. With a flexible interface, it uses three methods for glycan building. The first method is manual building ("Carbohydrate Builder" button). It allows selection of monosaccharide, as well as defining linkages, branching and substitution (Figure 9, middle). The second method involves the use of a template library (using "Oligosaccharide libraries" button) containing commonly relevant structures (http://glycam.org/Pre-builtLibraries.jsp). The third option (direct input from a text sequence) becomes relevant when the glycan structure does not exist in the library or challenging to build due to structural complexity. In this case, a text for the oligosaccharide in GLYCAM-Web's condensed notation can be entered as an input to create the glycan structure. Once the glycan is generated, the options include the solvation of the structure and the manual input of the glycosidic linkages. The tool allows structure minimisation and generates rotamers which can be visualised using JSmol viewer. Information about the force field that is used to build the structure is also provided. The multiple structures can be downloaded compressed as .tar, .gz or .zip files containing .pdb files. Similarly, the 2D image can be saved in GIF format. GLYCAM-web- Carbohydrate Builder can be used to prepare the system for MD simulation as it solvates the glycans and also generates the topology and coordinate files. In addition to its carbohydrate builder, Glycam-web consists of additional tools like glycoprotein builder and glycosaminoglycans (GAG) builder.

**CHARMM-GUI (Glycan Reader and Modeler).** The CHARMM-GUI (http://www.charmm-gui.org) is a web-based graphical user interface which provides various functional modules to prepare complex biomolecular systems and input files for molecular simulations. Glycan Reader and Modeler [65-67] is a part of CHARMM-GUI (Figure 9, bottom) and available as a freely accessible online tool at http://charmm-gui.org/input/glycan. It can read input files in PDB, PDBx/mmCIF and CHARMM formats containing glycans and automatically detects the carbohydrate molecules and glycosidic linkage information. Alternatively, it can also read a glycan sequence (GRS format) to generate a 3D model and input files for MD simulation of the carbohydrate-only system. GRS carbohydrate sequences can be made through a straightforward interface: monosaccharides (20+ options) and their linkages are added incrementally from drop-down menus. A useful feature of this tool is the real-time rendering of the carbohydrate image: each added monosaccharide and modified linkage is directly reported to the image as well as to a text (GRS) format. Option for numerous chemical modifications is also available.

On the other hand, the Glycan Modeler allows in silico N-/O-glycosylation for glycan-protein complexes and generates a "most relevant" glycan structure through Glycan Fragment Database (GFDB) [68] search which gives proper orientations relative to the target protein. In the absence of target glycan sequence in GFDB, the structures are generated by using the valid internal coordinate information (averaged phi, psi, and omega glycosidic torsion angles) in the CHARMM force field. Input files for CHARMM can be generated for the purpose of MD simulation. Amongst other possible outputs, 3D representations of the glycans are available as .pdb files. This tool can be helpful for researchers to generate 2D depictions of a glycan and then obtain the corresponding 3D representation, which can be useful for modelling studies of glycans and glycoconjugates.

**doGlycans.** doGlycans [69] is a compilation of tools designed for preparing carbohydrate structures for atomistic simulations of glycoproteins, carbohydrate polymers and glycolipids using GROMACS [70,71] In the form of Python scripts; the tools are used to prepare the system, which generally includes the processing of a .pdb file using the *pdb2gmx* tool. Subsequently, a glycosylation model can be prepared for carbohydrate polymer simulation using the *prepreader.py* script. Similarly, the *doglycans.py* script can be used to develop models for glycoproteins and glycolipids. Together, these tools are called doGlycans toolset. Although doGlycans is highly flexible, it only uses the sugar units that are defined in GLYCAM. The topologies generated for glycosylated proteins and glycolipids are compatible with the OPLS [72] and AMBER [73] force fields. The topology for carbohydrate polymers is based on the GLYCAM force field. The user needs to provide the ceramide topology as input to generate the topologies for glycolipids. The tools contained in doGlycans create 3D models and simulation files as a starting point for more complex molecular simulation studies.

**Figure 9:** From top to bottom: Sweet II [62] web-interface with a text input to generate a 3D model. GLYCAM Carbohydrate Builder [63] interface which accepts a text input for glycans and generates 3D models. CHARMM-GUI (Glycan reader and Modeler) [64] interface with a 3D structure output generated using a glycan sequence as input.

**RosettaCarbohydrate.** Rosetta is a software suite for macromolecular modelling as an extensive collection of computer code mostly written in C++ and Python languages. Rosetta is available to academic and commercial researchers through a license available at https://www.rosettacommons.org/software/license-and-download. The licence is free for academic users. The tool runs best on Linux or macOS platforms only. It can be installed on a multiprocessor computing cluster to increase efficiency. RosettaCarbohydrate [74,75] tool provides the methods for general modelling and docking applications for glycans and glycoconjugates. The application accepts the standard PDB, GLYCAM, and GlycoWorkbench (.gws) file formats and the available utilities (codes) helps with the general problems in sampling, scoring, and nomenclature related to glycan modelling. It samples glycosidic bonds, ring forms, side-chain conformations, and utilises a glycan-specific term within its scoring function. The tool also consists of utilities for virtual glycosylation, protein–glyco-ligand docking, and glycan "loop" modelling. This tool is best for the researcher with basic knowledge and skills to work with a command-line interface (Linux).

**PolysGlycanBuilder.** PolysGlycanBuilder [76] is a web-based tool (http://glycan-builder.cermav.cnrs.fr/) with an interactive and more usable interface (Figure 10). The software translates a glycan sequence or polysaccharide repeat unit into the coordi-nate set of the corresponding tertiary structure, in one or several of its low energy conformations. The construction follows an intuitive scheme which is as close as possible to the way glyco-scientists draw the sequence of their structures. The simplest method for model building involves dragging and dropping monosaccharide units to the canvas or workspace grid. The software displays rows of monosaccharides in the form of standard SNFG symbols with 3D information (furanose/pyranose shape, configuration, anomericity, and ring conformation). Glycosidic linkages can be easily defined, as the values of the dihedral angles ($\Phi$, $\Psi$, $\Omega$). They can be manually set or extracted from a database of low energy conformations of 600 disaccharide segments. The monosaccharides have been subjected to geometry optimisation using molecular mechanics approach. For a given input sequence, the corresponding 3D coordinates are generated at the PDB format. Within the process of construction, the structure is displayed via the LiteMol and eventually optimised to remove any steric clashes. The image for the glycan can be downloaded and saved in SVG format. Keeping the glycan/polysaccharide structure in text format (condensed IUAPC, GlycoCT, SNFG and INP) offers several ways to connect to other applications. Other than drag and drop method, PolysGlycan-Builder also accepts input of files in INP, IUPAC and GlycoCT formats. An interactive interface accompanies the application, which makes it more versatile for glycan drawing and 3D model building.

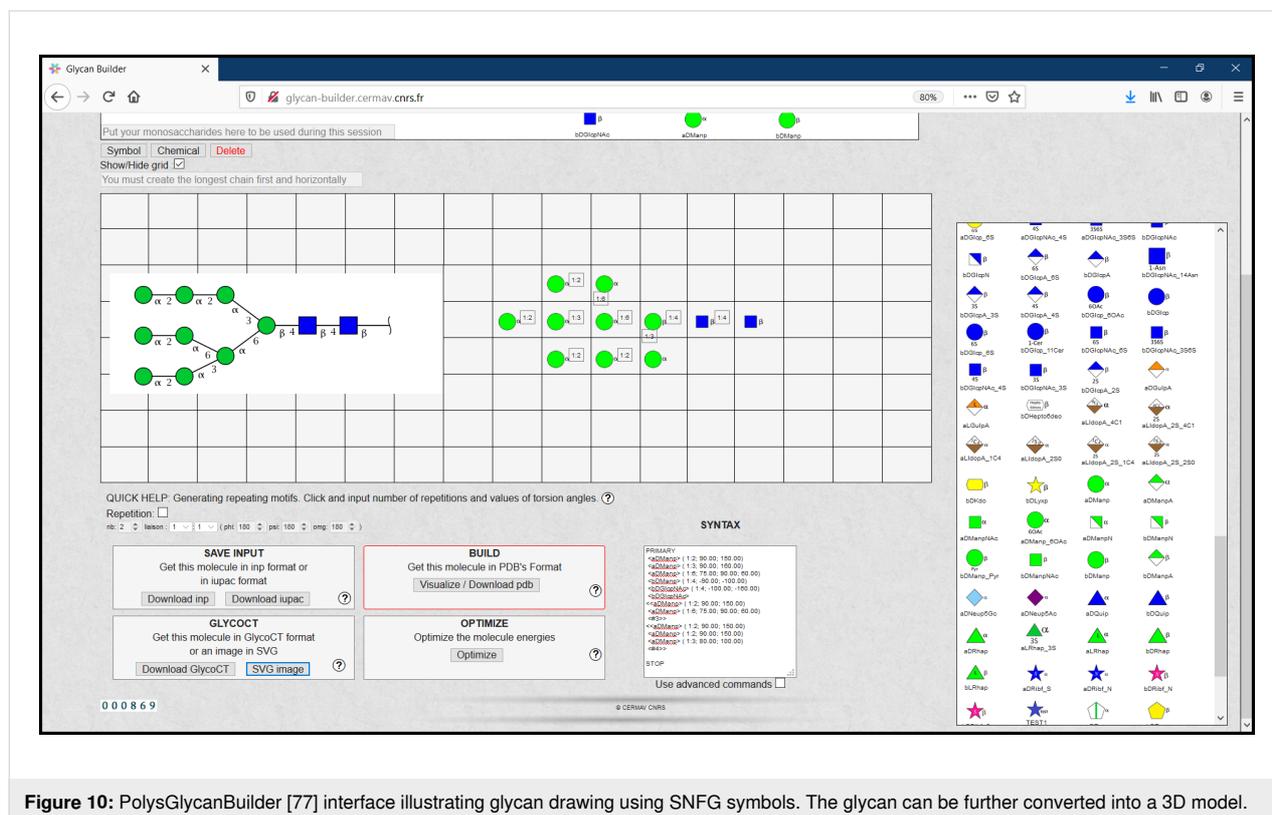

**Figure 10:** PolysGlycanBuilder [77] interface illustrating glycan drawing using SNFG symbols. The glycan can be further converted into a 3D model.

## Displaying 3D structures of glycans

**3D-SNFG VMD interface and visualisation algorithms.** The recently introduced 3D-Symbol Nomenclature for Glycans (3D-SNFG) [15] allows the representation of carbohydrates in an unusual way: the SNFG symbols are added to a three-dimensional structure. The 3D-SNFG script must be integrated into the visual molecular dynamics (VMD) [21,78] viewer software to enable the representation of glycans as large SNFG-matching 3D shapes that can either replace the molecular monosaccharides or stay lodged at the geometric centre of the cycle (Figure 11, top left). Upon the input of a glycan-containing structure (in PDB format), the integrated script in VMD automatically recognises the common monosaccharide names and generates the 3D shapes. The embedded script also enables shortcuts keys from keyboard to quickly change between large and small 3D-SNFG shapes and also label the reducing terminus. The 3D structure displayed in VMD can be saved as a .bmp image file. Thanks to 3D-SNFG, the standardised representation of glycan structures can finally take a step into the 3D space. The obtained images can become very useful for quick assessment of 3D glycan models.

In addition to the 3D-SNFG script, *PaperChain* and *Twister* [83] are two visualisation algorithms available with the Visual Molecular Dynamics (VMD) package. These algorithms are useful to visualize complex cyclic molecules and multi-branched polysaccharides.{Cross, 2009 #69} *PaperChain* displays rings in a molecular structure with a polygon and colours them according to the ring pucker. The other algorithm (*Twister*) traces glycosidic bonds in a ribbon representation that twists and changes its orientation according to the relative position of following sugar residues, hence provides an important conformational detail in polysaccharides. Combination of these algorithms with other visualisation features available in VMD can enhance the flexibility of displaying structural details of glycoconjugate, glycoprotein and cyclic structures.

**LiteMol.** The LiteMol [22] viewer is a freely available web application (Figure 11, top right) for 3D visualisation of macromolecules and other related data. LiteMol enables standard visualisation of macromolecules in different representation modes like surface, cartoons, ball-and-stick, etc. The software can be accessed at v.litemol.org and also available for integration in a webpage from the github (https://github.com/dsehnal/LiteMol). LiteMol is compatible with all modern browsers without the support of additional plugins. The viewer automatically depicts any carbohydrate residues and displays 3D structures of carbohydrates with 3D-SNFG symbols, which allows the viewer to identify the monosaccharides readily. The presented structure can be saved as a .png image file. Any monosaccharide with a residue name in PDB can be visualised using 3D-SNFG in LiteMol. However, a significant portion of the carbohydrates may contain some form of error in annotation, which would result in either no symbol or an incorrect symbol. Although LiteMol is an efficient and rapid 3D viewer for glycans, 3D representation does not provide any information about the glycosidic linkage type (e.g. α1-3 or β1-4). Also, it does not display any information about connection and configuration. If this information is required, returning to the classic molecular representation is possible.

**PyMOL- Azahar plugin.** Azahar [84] is a plugin in PyMOL [85] which enables building, visualization and analysis of glycans and glycoconjugates. This tool is based on Python and provides additional computing environment within the PyMOL package. The tool is provided with a template list of saccharide structures to facilitate structure building and visualisation. The interface provides three option menus to assist glycan structure building. The two first options help to specify residues to be connected from a list of available templates, and the third one allows selection of the chemical bond between the residues. The visualisation using PyMOL includes three cartoon-like representations. These display modes provided in the tool simplify the representation of glycan structures in cartoon, wire and bead representations. In cartoon and wire representations, the rings in sugars are shown as non-flat polygons connected by rods while in the bead representation mode, these cycles are represented as a sphere. In addition of visualization of static structures, the tool also allows analysis of trajectories of MD simulations. The tool can be used for conformational search using a Monte Carlo approach [86]. The conformational search is done by perturbing a torsional angle, followed by an energy minimization using the MMFF94 force field. Azahar is freely accessible from http://www.pymolwiki.org/index.php/Azahar.

**UnityMol/SweetUnityMol.** Sweet UnityMol [32] is a molecular structure viewer (Figure 11, middle) developed from the game engine Unity3D. The software is available for free download (https://sourceforge.net/projects/unitymol/files/UnityMol_1.0.37/) from the SourceForge project website. It can be installed in Mac, Windows and Linux platforms. The program reads files in PDB, mmCIF, Mol2, GRO, XYZ, and SDF formats, OpenDX potential maps and XTC trajectory files. It efficiently displays specific structural features for the simplest to the most complex carbohydrate-containing biomolecules. Sweet UnityMol displays 3D carbohydrate structures with different modes of representation, such as: liquorice, ball-and-stick, hyperBalls, RingBlending, hydrophilic/hydrophobic character of sugar face etc. The most recent version is fully compatible with the SNFG colour coding, which also uses acceptable pictorial representation, generally used in carbohydrate chemistry, biochemistry and glycobiology.

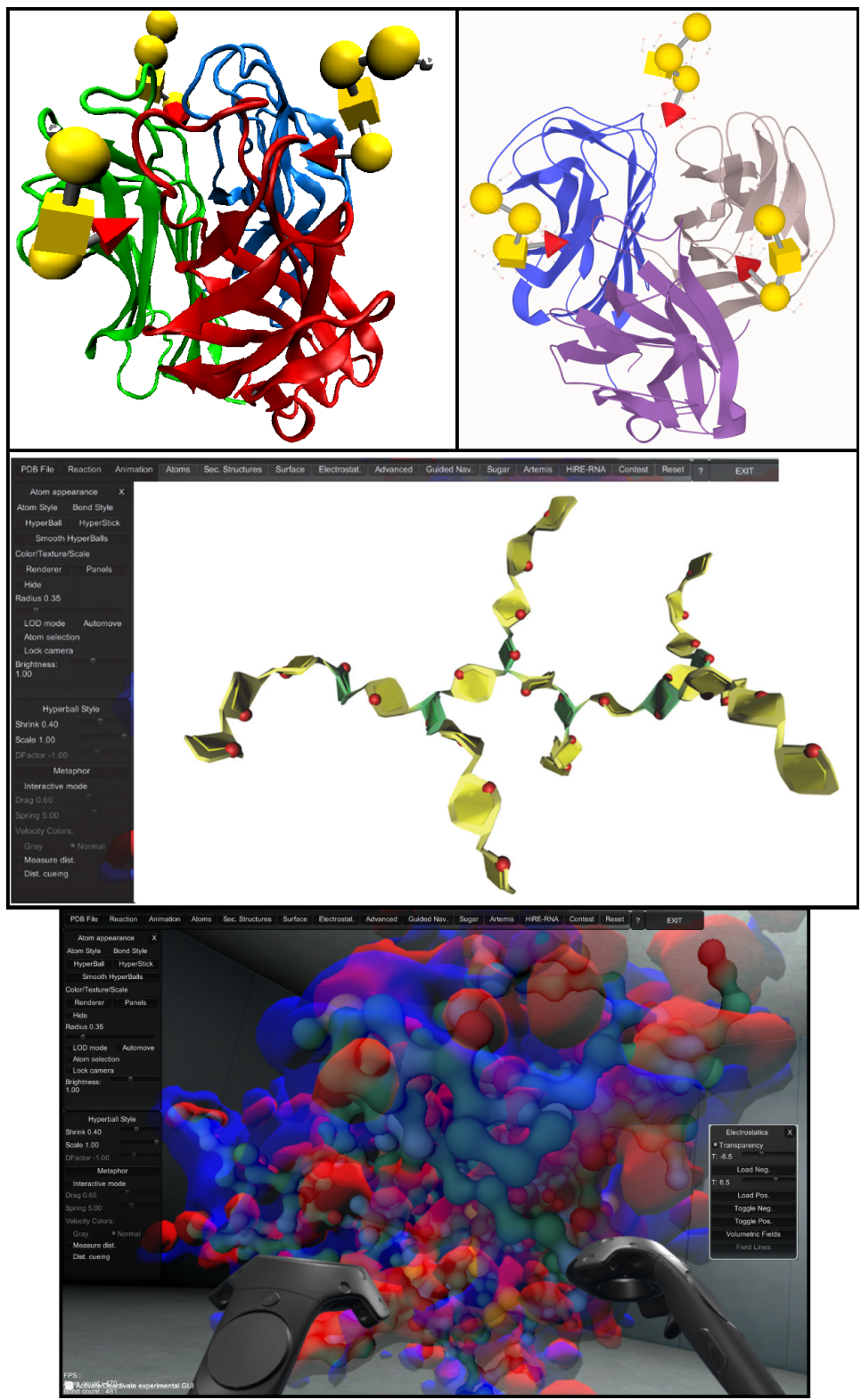

**Figure 11:** From top to bottom: 3D-SNFG representation of glycan using 3D-SNFG script integrated VMD [79]. LiteMol [80] interface with 3D-SNFG representation of glycan in a protein–glycan complex. SweetUnityMol [81] among the several types of representations a ribbon-like display of polysaccharide ribbons maintains the SNFG colour coding of monosaccharides. UnityMol [82] within an immersive virtual reality context.

SweetUnityMol provides a continuum from the conventional ways to depict the primary structures of complex carbohydrates all the way to visualising their 3D structures. Several options are offered to the user to select the most relevant type of depictions, including new features, such as "Coarse-Grain" representation while keeping the option to display the details of the atomic representations. Powerful rendering methods produce high-quality images of molecular structures, bio-macromolecular surfaces and molecular interactions.

A recently developed version of UnityMol has been implemented with the immersive Virtual Reality context using head-mounted displays [87]. It offers high-quality visual representations, ease of interactions with multiple molecular objects, powerful tools for visual manipulations, accompanied by the evaluation of intermolecular interactions. Consequently, simultaneous investigations of multiple objects such as macromolecular interactions gain in efficiency and accuracy. (Figure 11, bottom).

## Conclusion
The set of computational tools presented above illustrates the rich contributions of a community devoted to enabling the accurate representation of complex carbohydrates via the development and implementation of a versatile informatics toolbox. These legitimate efforts aim at facilitating communication within the scientific community. To establish a comparative analysis of the several available applications, we evaluated 17 selected items that characterise best their availability, implementation, maintenance and field of use. The comparative analysis of tools could be useful for glycobiologists or any researcher looking for a ready to use, simple application for the sketching, building and display of glycans.

This article provides an overview of the computational tools and resources available for glycan sketching, building and representing. It also provides a descriptive analysis of the recently developed software tools dedicated explicitly to glycans and glycoconjugates. The newly developed tools are more advanced and use the standard nomenclature and symbols for glycan representation. These tools can further help to standardise the description of glycans in research, communication and databases.

## Supporting Information

### Supporting Information File 1
Features of glycan sketchers, builders and viewers.
[https://www.beilstein-journals.org/bjoc/content/supplementary/1860-5397-16-199-S1.pdf]

## Acknowledgements
Appreciation is extended to Drs. A. Imberty, A. Varrot, L. Belvisi and A. Bernardi for their support.

## Funding
This research was performed within the framework of the PhD4GlycoDrug Innovative Training Network and was funded from the European Union's Horizon 2020 research and innovation programme under the Marie Skłodowska-Curie grant agreement No 765581. The work was supported by the Cross-Disciplinary Program Glyco@Alps, within the framework "Investissement d'Avenir" program [ANR-15IDEX-02].

## ORCID® iDs
Kanhaya Lal - https://orcid.org/0000-0001-8555-7948
Rafael Bermeo - https://orcid.org/0000-0002-4451-878X
Serge Perez - https://orcid.org/0000-0003-3464-5352

## License and Terms